\documentclass[final,authoryear,5p,times,twocolumn]{elsarticle}

\usepackage{natbib}

\usepackage{graphicx}

\journal{New Astronomy}

\usepackage{amssymb}

\def\astrobj#1{#1}

\begin{document}

\begin{frontmatter}



\title{Discovery of deep eclipses in the cataclysmic variable \astrobj{IPHAS J013031.89+622132.3}}

\author{V. P. Kozhevnikov\corref{cor1}},

\ead{valery.kozhevnikov@urfu.ru}

\cortext[cor1]{Tel.: + 7 343 2615431; fax: + 7 343 3507401.}

\address{Astronomical Observatory, Ural Federal University, Lenin Av. 51,
Ekaterinburg, 620083, Russia}

\begin{abstract}

I performed photometric observations of the poorly studied cataclysmic variable \astrobj{IPHAS~J013031.89+622132.3} and discovered very deep eclipses. I obtained observations over 14 nights for a total time of 50 hours during a time span of 6 months. Thanks to the long observation interval, I determined the orbital period with high precision, $P_{\rm orb}=0.149\,350\,14\pm0.000\,000\,20$~d.  I derived the eclipse ephemeris, which, thanks to the precision of the orbital period, has a formal validity of 300 years. The average eclipse depth was $1.88\pm0.07$~mag. The prominent parts of the eclipses were smooth and symmetrical. The average eclipse width, including extended asymmetric eclipse wings, was $0.18\pm0.01$~phases or $40\pm2$~min. The average orbital light curve did not show a prominent orbital hump. Because no dwarf nova outburst occurred during the 6 months of monitoring, this cataclysmic variable is likely to be a nova-like variable. 

\end{abstract}


\end{frontmatter}

\section{Introduction}

Cataclysmic variables (CVs) are interacting binary stars consisting of a white dwarf and a late-type companion. The late-type companion fills its Roche lobe and transfers material to the white dwarf. Depending on the presence of large outbursts, CVs are subdivided into dwarf novae and nova-like variables. Classical and recurrent novae except for symbiotic stars are also reckoned among CVs. Such novae in quiescence are similar to nova-like variables. In non-magnetic systems, accretion occurs through accretion discs. In nova-like variables, accretion discs are permanently very bright. In dwarf novae, accretion discs are moderately bright during quiescence and very bright during outburst. Accretion discs often reveal bright spots, which form at the place where the accretion stream impacts the disc. These bright spots cause orbital humps in orbital light curves. These humps are more prominent in dwarf novae during quiescence and are less noticeable in nova-like variables. Comprehensive reviews of CVs are given in \citet{ladous94}, \citet{warner95} and \citet{hellier01}.

If the orbital inclination is high ($i > 60^\circ$), CVs show eclipses (e.g., \citealt{ladous94}). Eclipsing CVs are important for several reasons. First, eclipses make it possible to accurately determine the orbital period and then study its change.  Second, eclipses make it possible to reliably determine the orbital inclination, which is necessary to determine the masses of stellar components using radial velocity measurements (e.g., \citealt{hellier01}). Third, eclipses of a white dwarf and eclipses of a bright spot distinguishable in light curves make it possible to determine the masses of stellar components using only photometric data. (e.g., \citealt{zorotovic11}). Finally, eclipses allow one to study the structure and time evolution of accretion discs using eclipse mapping methods (e.g., \citealt{baptista04}).

Using the Isaac Newton Telescope Photometric H$\alpha$ Survey (IPHAS), \citet{witham07} discovered 11 new CV candidates. \citeauthor{witham07} performed follow-up observations of only three of them. To find the orbital periods, \citeauthor{witham07} performed spectroscopic observations of two CV candidates, \astrobj{IPHAS~J013031.89+622132.3} and \astrobj{051814.33+294113.0}. For both stars, these observations gave ambiguous results due to the aliasing problem (see Fig.~6 in \citeauthor{witham07}). Photometric observations of these two CV candidates were not performed. In contrast, the third CV candidate, \astrobj{IPHAS~J062746.41+014811.3}, was observed by \citeauthor{witham07} only photometrically.  They discovered this star to be an eclipsing CV. Later, \citet{aungwerojwit12} refined this result by discovering that \astrobj{IPHAS~J062746.41+014811.3} is an eclipsing intermediate polar with an eclipse depth of 1.3~mag. Moreover, \citeauthor{witham07} highlighted another system, \astrobj{IPHAS~J025827.88+635234.9}, which might be a high-luminosity object reminiscent of \astrobj{V~Sge}. Detailed photometric observations of this bright (13.5 mag) CV candidate were not performed. Following the suggestion by \citeauthor{witham07} that \astrobj{IPHAS~J025827.88+635234.9} may be a member of the V~Sge class, I observed it photometrically and ruled out this possibility. Instead, I found that \astrobj{IPHAS~J025827.88+635234.9} is an ordinary eclipsing CV with an eclipse depth of 0.3~mag \citep{kozhevnikov14}. Lately, I improved my photometric technique. This allowed me to observe faint stars like \astrobj{IPHAS~J013031.89+622132.3} and \astrobj{051814.33+294113.0}.  In \citet{kozhevnikov18}, I continued the study of these CV candidates and discovered deep eclipses (2.4 mag) in \astrobj{IPHAS~J051814.33+294113.0}. Then, I monitored \astrobj{IPHAS~J013031.89+622132.3}  (hereafter \astrobj{J0130}) and also discovered deep eclipses. In this paper I present the results obtained from these observations.

\section{Observations}

\begin{table}
\caption[ ]{Journal of the observations}
\label{journal}
\begin{flushleft}
\begin{tabular}{lcc}
\noalign{\smallskip}
\hline
\noalign{\smallskip}
Date                            & BJD$_{\rm TDB}$ start        & Length \\
(UT)                            & (-2,458,000)                           & (h) \\
\noalign{\smallskip}
\hline
\noalign{\smallskip}
February 16             & 166.126823                              & 10.2 \\
February 18             & 168.152645                              & 2.8 \\
August 12                & 343.278899                              & 3.2 \\
August 15                & 346.261779                              & 3.8 \\
August 17                & 348.260525                              & 4.0 \\
August 19                & 350.284742                              & 3.2 \\
August 20                & 351.321358                              & 0.8 \\
September 3            & 365.208981                              & 2.1 \\
September 4            & 366.223281                              & 4.3 \\
September 5            & 367.207106                             & 3.1 \\
September 11          & 373.223956                             & 4.3 \\
September 12          & 374.238767                             & 2.1 \\
September 13          & 375.316815                             & 3.9 \\
September 16          & 378.398499                             & 1.7 \\
\noalign{\smallskip}
\hline
\end{tabular}
\end{flushleft}
\end{table}


I performed the photometric observations of J0130 at Kourovka observatory, Ural Federal University, using the 70-cm Cassegrain telescope and the multi-channel pulse-counting photometer. The observatory is located at a distance of 80 km from Ekaterinburg. The design of the photometer is described in \citet{kozhevnikoviz}. Its three channels enable to measure the brightness of two stars and the sky background simultaneously. The mounting of the telescope is equipped with computer-controlled step motors. To maintain precise centring of the two stars in the photometer diaphragms, I use the CCD guiding system. This guiding system and computer-controlled step motors make it possible to perform brightness measurements automatically and nearly continuously. Short interruptions occur according to a computer program when I measure the sky background in all three channels simultaneously. This is necessary to define the differences in the sky background, which are caused by differences in the size of the diaphragms, and to eliminate the effect of faint stars in the sky background channel.

Although the multi-channel photometer provides high-quality photometric data even under unfavourable atmospheric conditions \citep[e.g.,][]{kozhevnikov02}, for a long time I could not observe stars fainter than 15~mag that are invisible to the eye. A few years ago, I realized that, with the aid of step motors of the telescope, I can centre two stars in the diaphragms, one of which is invisible, using the coordinates of the invisible star and the coordinates of a nearby reference star. Using this method, I can observe very faint stars up to 20~mag \citep[e.g.,][]{kozhevnikov18}. 

Photometric observations of \astrobj{J0130} were performed in February--September 2018 over 14 nights with a total duration of 50~h. The data were obtained in white light (approximately 3000--8000~\AA). The time resolution was 16 s. For \astrobj{J0130} and the comparison star, I used diaphragms of $16''$. For the sky background, I used a diaphragm of $30''$. The comparison star is \astrobj{USNO-A2.0 1500-01531296}. It has $B=14.1$~mag and $B-R=0.9$~mag. The colour index of \astrobj{J0130} is similar to the colour index of this star.  According to the USNO-A2.0 catalogue, \astrobj{J0130} has $B=17.5$~mag and $B-R=0.6$~mag.  Similar colour indexes of these two stars reduce the effect of differential extinction. 

I obtained differences of magnitudes of \astrobj{J0130} and the comparison star by taking into account differences in light sensitivity between different channels. Using the comparison of the sky background counts in the sky background channel, the sky background counts were corrected for the effect of faint stars, which may accidentally be in the sky background diaphragm. To measure the sky background in stellar diaphragms, I carefully found a place without faint stars using star maps provided by Astronet (http://www.astronet.ru/db/map). These star maps are based on USNO-A2.0 $B$ magnitudes. However, this did not allow me to avoid a faint star in the diaphragm of \astrobj{J0130} when I measured the sky background. The counts in the diaphragm of \astrobj{J0130}, from which the sky background counts were subtracted, were mainly negative near the mid-eclipses. Unfortunately, I noticed this during the final data processing.

Using Aladin Lite (https://aladin.u-strasbg.fr/AladinLite) and VizieR (http://vizier.u-strasbg.fr/viz-bin/VizieR), I identified this faint star. It is \astrobj{Gaia DR2 510854650123377152} with $G=19.05$~mag \citep{gaia16, gaia18}. During discrete measurements of the sky background, this faint star was located in the diaphragm of \astrobj{J0130} near its edge. The spectral response of the photomultipliers (S20 photocathodes) suggests that my white light magnitudes are close to Gaia $G$ magnitudes (e.g., \citealt{maiz18}). Using the G magnitude of this faint star, I estimated that the sky background counts in the diaphragm of \astrobj{J0130} should be reduced by 1.5--2.8\%. After this correction, most of the counts in the diaphragm of \astrobj{J0130} near the mid-eclipses were positive. To evaluate the precision of this correction, I assumed that the faint star was 0.2~mag fainter.  This can be caused by poor centring of this star or by a discrepancy between my white light magnitudes and the $G$ magnitudes.  Then the average eclipse was 0.15~mag deeper. 

A journal of the observations is shown in Table~\ref{journal}. This table gives BJD$_{\rm TDB}$, which is the  Barycentric Julian Date in the Barycentric Dynamical Time (TDB) standard. BJD$_{\rm TDB}$ is uniform and therefore preferred. I calculated BJD$_{\rm TDB}$ using the online calculator (http://astroutils.astronomy.ohio-state.edu/time/) \citep{eastman10}. In addition, I calculated BJD$_{\rm UTC}$. During my observations, the difference between BJD$_{\rm TDB}$ and BJD$_{\rm UTC}$ was constant. BJD$_{\rm TDB}$ exceeded BJD$_{\rm UTC}$ by  69~s.

\begin{figure}[t]
\resizebox{\hsize}{!}{\includegraphics{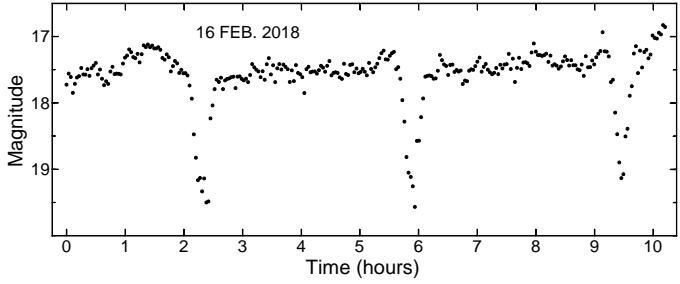}}
\caption{Longest light curve of \astrobj{J0130}, which shows three consecutive eclipses.}
\label{figure1}
\end{figure}

\section{Analysis and results}

\begin{table}
\caption[ ]{Parameters of the observed eclipses}
\small
\label{eclipses}
\begin{flushleft}
\begin{tabular}{lccc}
\noalign{\smallskip}
\hline
\noalign{\smallskip}
Date    & BJD$_{\rm TDB}$ mid-ecl.  &  Eclipse depth  & Out-of-ecl.\\
(UT)     & (-2,458,000)                         &   (mag)            &  magnitude \\
\noalign{\smallskip}
\hline
\noalign{\smallskip}
February 16     &  166.22433(27)                 & $1.95\pm0.10$   &  17.5\\
February 16     &  166.37327(22)                 & $1.95\pm0.08$   &  17.5  \\
February 16     &  166.52195(21)                 & $1.84\pm0.09$   &  17.5 \\
February 18     &  168.16491(19)                 & $2.23\pm0.08$   &  17.4 \\
August 12        &  343.35331(22)                 & $1.97\pm0.08$    & 17.1 \\
August 15        &  346.33969(21)                 & $1.66\pm0.07$   &  17.3  \\
August 17        &  348.28208(19)                 & $1.44\pm0.05$   &  17.3 \\
August 19        &  350.37213(16)                 & $2.18\pm0.07$   &  17.6  \\
September 4    &  366.35288(24)                 & $1.63\pm0.08$   &  17.6   \\
September 5    &  367.24909(25)                 & $2.16\pm0.10$   &  17.4  \\
September 11  &  373.37170(30)                 & $1.63\pm0.09$   &  17.7  \\
September 12  &  374.26806(15)                 & $2.24\pm0.07$   &  17.6  \\
September 13  &  375.46400(24)                 & $1.60\pm0.06$   &  17.7  \\
September 16  &  378.45038(36)                 & $1.81\pm0.12$   &  17.6 \\
\noalign{\smallskip}
\hline
\end{tabular}
\end{flushleft}
\end{table}

Fig.~\ref{figure1} shows the longest light curve of \astrobj{J0130}, in which three consecutive eclipses are obvious. For usability, the differential magnitudes were converted into the magnitudes using the average of the $B$ and $R$ magnitudes of the comparison star. To reduce the photon noise, counts were previously averaged over 128-s time intervals. The photon noise of the out-of-eclipse light curve (rms) is 0.06~mag. The photon noise near the mid-eclipses is undefined. It is obvious that the accuracy of the points near the mid-eclipses depend mainly on the accuracy of subtracting the sky background because the sky background near the mid-eclipses considerably exceeds the star counts. None the less, as seen in Fig.~\ref{figure1}, the eclipses have roughly equal depths.

During observations of \astrobj{J0130}, I obtained 14 whole eclipses. Table~\ref{eclipses} shows their parameters. The out-of-eclipse magnitudes were obtained by averaging of out-of-eclipse light curves. The mid-eclipse times and the eclipse depths were measured using a Gaussian function fitted to the eclipse and two adjacent parts of the out-of-eclipse light curve. The average eclipse depth is $1.88\pm0.07$~mag. As shown above, if I underestimated the sky background excess caused by the faint star, which accidentally was in the diaphragm of \astrobj{J0130} when I measured the sky background, the eclipse depth may be several tenths of a magnitude larger.  In contrast, if I overestimated the sky background excess, the eclipse depth may be several tenths of a magnitude less. However, in any case, the eclipses in \astrobj{J0130} seem very deep. 

It is well known that a direct fit of an ephemeris to a series of mid-eclipse times gives the best precision of the eclipse period. Obviously, this method is good when eclipses are evenly distributed in time. This is not consistent with my observations of \astrobj{J0130}. Therefore, to find the eclipse period, I use various methods and then compare the results. 

Three consecutive eclipses (Fig.~\ref{figure1}) allow me to find the approximate eclipse period. These measurements are useful because they eliminate the aliasing problem, which might arise as a result of complicated analysis. Using the three mid-eclipse times, which are presented in Table~\ref{eclipses}, I obtained $P_{1}=0.14894\pm0.00035$~d and $P_{2}=0.14869\pm0.00030$~d. The average value is $0.14882\pm0.00024$~d.

\begin{figure}[t]
\resizebox{\hsize}{!}{\includegraphics{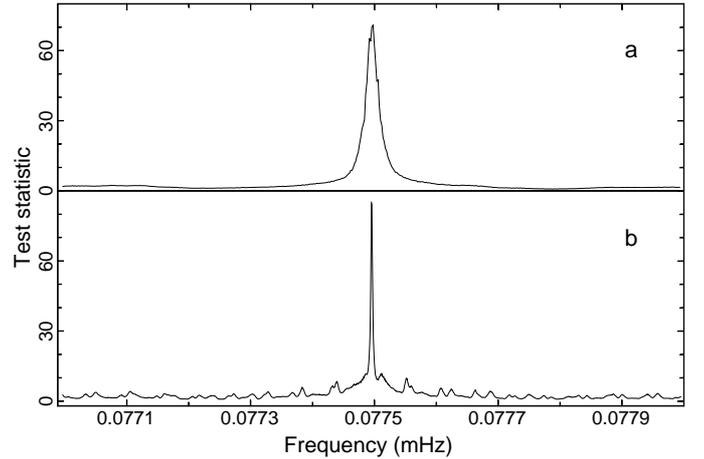}}
\caption{Analysis of Variance spectra of \astrobj{J0130} calculated for the data obtained in August--September, which contain 10 eclipses (a), and for all data, which additionally contain 4 eclipses observed in February (b).}
\label{figure2}
\end{figure}
 
To improve the precision of the eclipse period, I used the Analysis of Variance (AoV) method \citep{schwarzenberg89}, which is advantageous in comparison with the Fourier transform for non-sinusoidal signals \citep{schwarzenberg98}.  At first, I used only the data obtained in August--September because they are densely distributed in time. These data contain 10 whole eclipses. Fig~\ref{figure2}a shows the AoV spectrum of these data. This AoV spectrum reveals the principal peak with no aliases. The precise peak maximum was found using a Gaussian function fitted to the upper part of the peak.  The error was defined according to the method of \citet{schwarzenberg91}. This method considers both frequency resolution and noise and is appropriate for Fourier power spectra. For AoV spectra, I use it as a tentative one. From this AoV spectrum, the eclipse period is $0.149\,3504\pm0.000\,0029$~d.

Having ascertained that the AoV spectrum of the data of August--September showed no aliases, I calculated the AoV spectrum of all data, which additionally contain 4 eclipses observed in February. As seen in Fig.~\ref{figure2}b, the AoV spectrum of all data reveals a much narrower principal peak. This AoV spectrum also shows no aliases. From the AoV spectrum presented in Fig.~\ref{figure2}b, the eclipse period is $0.149\,350\,29\pm0.000\,000\,58$~d. The period precision obtained from all data is 5 times better than the period precision obtained from the data of August--September.

The precision of the eclipse period found from the AoV spectra is sufficient to avoid ambiguity in determining the number of cycles. Therefore I determined the eclipse period using independent pairs of distant eclipses. These four eclipse pairs and four corresponding eclipse periods are presented in Table~\ref{pairs}. The average period is $0.149\,350\,23\pm0.000\,000\,25$~d. The period error is approximately two times less than the error obtained from the AoV spectrum of all data according to \citet{schwarzenberg91}. Hence, the error found from the AoV spectrum is at least two times overestimated. 

\begin{table}
\caption[ ]{Period determination from pairs of distant eclipses}
\scriptsize
\label{pairs}
\begin{flushleft}
\begin{tabular}{lccc}
\noalign{\smallskip}
\hline
\noalign{\smallskip}
BJD$_{\rm TDB}$ mid-ecl. 1 & BJD$_{\rm TDB}$ mid-ecl. 2 &  Number     & Period \\
(-2,458,000)                           &  (-2,458,000)                         &   of cycles    & (days) \\
\noalign{\smallskip}
\hline
\noalign{\smallskip}
166.22433(27)     &    348.28208(19)    &    1219   & 0.14935008(27)      \\
166.37327(22)     &    350.37213(16)    &    1232   & 0.14934973(22)      \\
166.52195(21)      &   374.26806(15)    &    1391    & 0.14935018(18)      \\
168.16491(19)      &    375.46400(24)   &    1388    & 0.14935093(22)      \\
\hline
\noalign{\smallskip}
& & Average period:      &    0.14935023(25)   \\
\noalign{\smallskip}
\hline
\end{tabular}
\end{flushleft}
\end{table}

\begin{table}
\caption[ ]{Verification of the eclipse ephemeris}
\small
\label{ephemeris}
\begin{flushleft}
\begin{tabular}{lccc}
\noalign{\smallskip}
\hline
\noalign{\smallskip}
Date & BJD$_{\rm TDB}$ mid-ecl.  & Number          &  O--C $\times 10^{3}$ \\
(UT) & (-2,458,000)                         & of cycles        &  (days)               \\
\noalign{\smallskip}
\hline
\noalign{\smallskip}
February 16  & 166.22433(27)       &      -1393       & $0.60\pm0.31$      \\
February 16  & 166.37327(22)       &      -1392      & $0.19\pm0.27$      \\
February 16  & 166.52195(21)       &      -1391      & $-0.47\pm0.26$      \\
February 18  & 168.16491(19)       &      -1380      & $-0.37\pm0.25$      \\
August 12     & 343.35331(22)       &      -207        & $0.32\pm0.27$       \\
August 15     & 346.33969(21)       &      -187        & $-0.31\pm0.26$       \\
August 17     & 348.28208(19)       &      -174        & $0.53\pm0.25$        \\
August 19     & 350.37213(16)       &      -160        & $-0.32\pm0.22$       \\
September 4 & 366.35288(24)       &       -53         & $-0.03\pm0.28$       \\
September 5 & 367.24909(25)       &       -47         & $0.08\pm0.30$        \\
September 11 & 373.37170(30)      &        -6         & $-0.67\pm0.33$       \\
September 12 & 374.26806(15)      &         0          & $-0.42\pm0.21$       \\
September 13 & 375.46400(24)      &         8          & $0.73\pm0.28$        \\
September 16 & 378.45038(36)      &         28        & $0.10\pm0.39$        \\
\noalign{\smallskip}
\hline
\end{tabular}
\end{flushleft}
\end{table}

\begin{figure}[t]
\resizebox{\hsize}{!}{\includegraphics{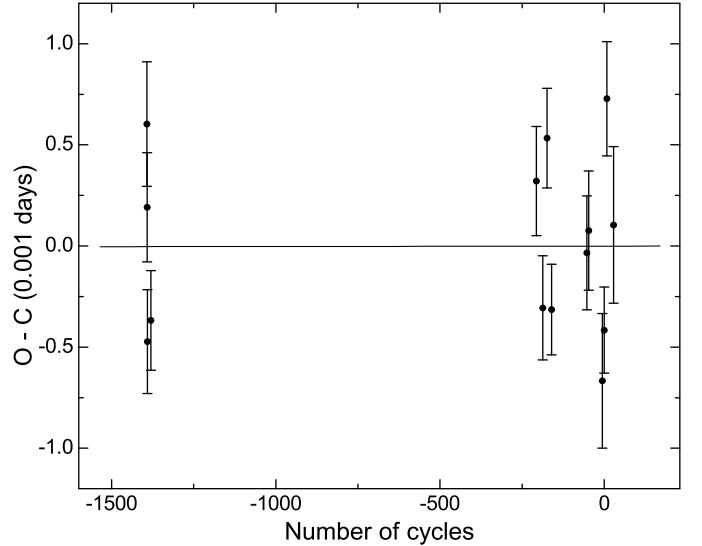}}
\caption{O -- C diagram calculated using the final eclipse ephemeris.}
\label{figure3}
\end{figure}

Using the most precise mid-eclipse time (see Table~\ref{eclipses}) and the most precise eclipse period obtained from pairs of distant eclipses, I obtained the following tentative ephemeris:

{\small
\begin{equation}
\small BJD_{\rm TDB}(\rm mid-ecl.)= 245\,8374.268\,06(15)+0.149\,350\,23(25) {\it E}.
\label{ephemeris1}
\end{equation} }

Using this ephemeris, I calculated O -- C for 14 eclipses. These O -- C obey the linear relation: 
O -- C = 0.000\,41(15) -- 0.000\,0009(20) {\it E}. It reveals a significant displacement along the vertical axis. I corrected the tentative ephemeris using the coefficients of this relation and obtained the final ephemeris:

 {\small
\begin{equation}
\small BJD_{\rm TDB}(\rm mid-ecl.)= 245\,8374.268\,47(15)+0.149\,350\,14(20) {\it E}.
\label{ephemeris2}
\end{equation} }

Using ephemeris~\ref{ephemeris2}, I calculated O -- C and presented them in Table~\ref{ephemeris} and in Fig.~\ref{figure3}. As seen, the O -- C diagram reveals no slope and displacement. The eclipse period and its rms error in ephemeris~\ref{ephemeris2} obtained from the linear fit of ephemeris~\ref{ephemeris1} to the series of mid-eclipse times, are close to the eclipse period and to its rms error obtained from pairs of distant eclipses. The time during which the accumulated error from the period runs up to one oscillation cycle is considered as a formal validity of an ephemeris. Based on the error of the eclipse period, the formal validity of ephemeris~\ref{ephemeris2} is 300 years (a confidence level of $1\sigma$). This means that ephemeris~\ref{ephemeris2} can be used without ambiguity in determining the number of cycles during 300 years. Table~\ref{periods} presents the eclipse periods obtained by different methods. All periods are compatible with each other. 


\begin{figure*}[t]
\resizebox{\hsize}{!}{\includegraphics{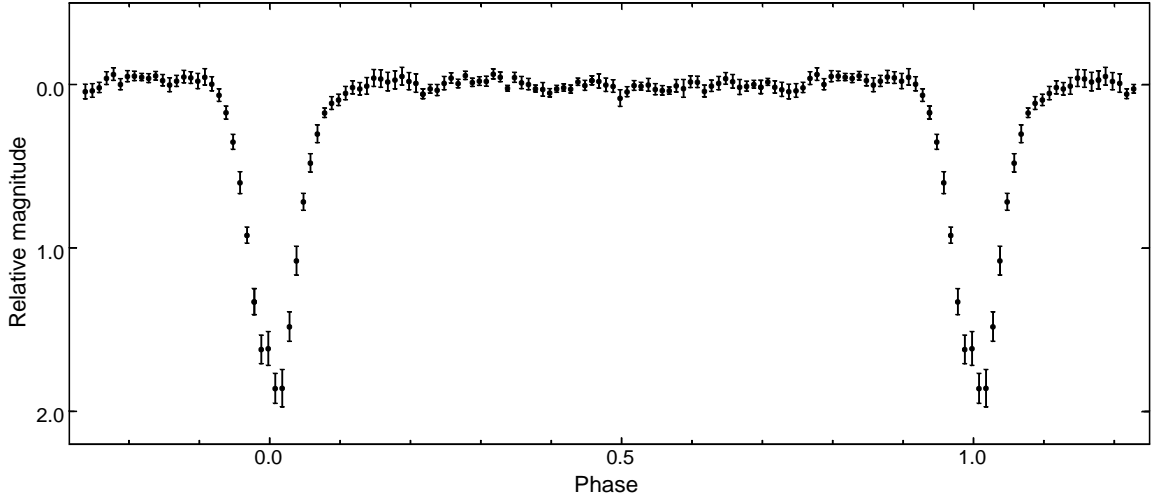}}
\caption{Light curves of \astrobj{J0130} folded with a period of 0.14935014~d. The depth of the average eclipse is $1.87\pm0.03$~mag.}
\label{figure4}
\end{figure*}

I folded 12 light curves containing 14 whole eclipses. As seen in Fig.~\ref{figure4}, the prominent parts of the eclipses are smooth and symmetrical, and the eclipses have asymmetric extended wings. The eclipse width including the extended eclipse wings is $0.18\pm0.01$~phases or $40\pm2$~min. From the folded light curve, I determined the depth of the average eclipse using a Gaussian function fitted to the average eclipse and two adjacent parts of the out-of-eclipse light curve. The depth of the average eclipse is $1.87\pm0.03$~mag. This depth is nearly the same as I obtained from direct averaging of the depths of individual eclipses. As seen in Fig.~\ref{figure4}, the out-of-eclipse light curve reveals no prominent orbital hump.

\begin{table}
\caption[ ]{Comparison of periods obtained by different methods}
\small
\label{periods}
\begin{flushleft}
\begin{tabular}{lll}
\noalign{\smallskip}
\hline
\noalign{\smallskip}
Method                                 & Period           &   Deviation from           \\
                                             & (days)           &   linear fit                      \\
\noalign{\smallskip}
\hline
\noalign{\smallskip}
Adjacent eclipses                &   0.14882(24)           &    $2.2\sigma$      \\
AoV of August--September &   0.1493504(29)       &    $0.1\sigma$     \\
AoV of all data                     &   0.14935029(58)     &    $0.3\sigma$     \\
Pairs of distant eclipses      &   0.14935023(25)      &    $0.2\sigma$    \\
Linear fit of the ephemeris  &   0.14935014(20)      &  --                        \\
\noalign{\smallskip}
\hline
\end{tabular}
\end{flushleft}
\end{table}

\begin{figure}[t]
\resizebox{\hsize}{!}{\includegraphics{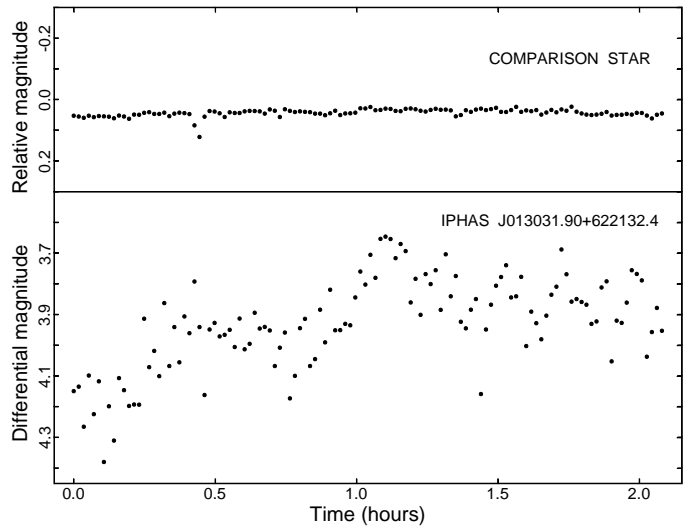}}
\caption{Segment of the light curve of \astrobj{J0130} obtained with a time resolution of 64~s in September 11, which shows noticeable flickering. The photon noise of the light curve (rms) is 0.1~mag.}
\label{figure5}
\end{figure}

All CVs reveal random brightness changes called flickering, which are visible directly in the light curves. Because \astrobj{J0130} is a faint star, such brightness changes are difficult to notice due to the large photon noise. None the less, I found a segment of the light curve, in which random brightness changes are clearly visible and can be attributed to flickering (Fig.~\ref{figure5}). From this segment, I estimated a flickering peak-to-peak amplitude of about 0.2~mag. 

To search for rapid coherent oscillations similar to oscillations in intermediate polars, I analysed the out-of-eclipse light curves obtained with a time resolution of 16~s. The careful Fourier analysis did not reveal rapid coherent oscillations exceeding the noise level. The maximum semi-amplitude of the noise peaks was 15~mmag at frequencies higher than 1.3~mHz. At lower frequencies, the noise peaks had noticeably higher semi-amplitudes due to flickering. 

At low frequencies, I attempted to find peaks with periods close to the orbital period. Such peaks might be evidence of superhumps. The result was ambiguous. The Fourier power spectrum of the out-of-eclipse light curves obtained in August--September revealed a coherent oscillation with a semi-amplitude of 70~mmag and a period of $0.14440\pm0.00021$~d. This oscillation can be attributed to negative superhumps because its period is 3.3\% less than the eclipse period.  However, the August data and the September data analysed separately did not show this oscillation.

\section{Discussion}

I performed photometric observations of \astrobj{J0130} and detected deep eclipses for the first time. By analyzing the light curves contained 14 whole eclipses, I accurately measured the orbital period, $P_{\rm orb}=0.149\,350\,14\pm0.000\,000\,20$~d. The high precision in the period determination was obtained thanks to the large coverage of the observations, which is six months. As seen in Fig.~\ref{figure2}, my data are densely distributed in time to avoid the aliasing problem. The sharp eclipses were also favourable to avoid the aliasing problem. From spectroscopic observations, however, \citet{witham07} suggested a different orbital period, but their periodogram was heavily aliased. This led to a large error.


The reliably determined orbital period and the precise average orbital light curve (Fig.~\ref{figure4}) allow me to make the assumption about the subtype of \astrobj{J0130}. The long orbital period well above the period gap (measured only in 18\% of dwarf novae, see \citealt{ritter03}) and the absence of luminosity variations outside the eclipses are more typical of nova-like variables than of other types of CVs. In addition, during six months of observations, \astrobj{J0130} did not show large brightness changes resembling dwarf nova outbursts. Thus, \astrobj{J0130} is most probably a nova-like variable.

The average eclipse depth obtained from direct averaging of 14 individual eclipses (Table~\ref{eclipses}) is $1.88\pm0.07$~mag. This eclipse depth is nearly equal to the eclipse depth in the folded light curve (Fig.~\ref{figure4}), which is $1.87\pm0.03$~mag. The eclipses discovered in \astrobj{J0130} are very deep. Such deep eclipses are more typical of polars because polars are disc-less systems. But I have ruled out that this CV is a polar because it shows no significant brightness variations outside of eclipses (see, e.g., \citealt{warner95}). In contrast with polars, only 30\% of eclipsing nova-like variables and nova remnants show eclipses deeper than 1.8~mag (see \citealt{ritter03} and references therein). Their total number is 20. So, \astrobj{J0130} is one of seldom disc systems with deep eclipses. This characteristic makes it suitable for future work with eclipse mapping (see, e.g., \citealt{baptista04}).

Using the precise eclipse period, I derived the eclipse ephemeris with a formal validity of 300 years. This means that ephemeris~\ref{ephemeris2} allows one to calculate O -- C without ambiguity in determining the number of cycles during 300 years. Therefore, ephemeris~\ref{ephemeris2} is suitable for future investigations of long-term changes of the orbital period using O -- C. Such changes might be caused by variations of the rotational oblateness of the late-type companion during its activity cycle (e.g., \citealt{rubenstein91, applegate92}). Such changes might also be caused by a possible giant planet orbiting around the centre of gravity in \astrobj{J0130} (e.g., \citealt{bruch14, beuermann11}). 

Eclipses observed in binary stars make it possible to determine the orbital inclination and then determine the masses of stellar components from radial velocity measurements. In CVs, this task is complicated because the white dwarf spectrum is directly invisible and, in radial velocity measurements, is replaced by the accretion disc spectrum. Because this spectrum can be contaminated with asymmetric disc structures \citep{robinson92}, measurements of radial velocities are difficult. In future spectroscopic observations of \astrobj{J0130}, the knowledge of orbital phases defined using ephemeris~\ref{ephemeris2} can help to identify these structures and thus solve this task (e.g., \citealt{hellier01}). To determine the masses of stellar components in \astrobj{J0130}, radial velocity measurements can be used together with photometric data. Such measurements make it possible to simultaneously determine the masses of stellar components and the orbital inclination (e.g., \citealt{szkody93, downes86}).

\section{Conclusions}

I performed photometric observations of the cataclysmic variable \astrobj{J0130} and detected very deep eclipses for the first time. The analysis of these data gives the following results:

\begin{enumerate}

\item My prolonged monitoring allowed me to determine the orbital period in \astrobj{J0130} with high precision, $P_{\rm orb}=0.149\,350\,14\pm0.000\,000\,20$~d. 
\item The average eclipse depth was very large and equal to $1.88\pm0.07$~mag. 
\item The prominent parts of the eclipses were smooth and symmetrical. 
\item The eclipses revealed extended asymmetric wings.
\item The average eclipse width including extended eclipse wings was $0.18\pm0.01$~phases or $40\pm2$~min.
\item I derived the eclipse ephemeris, which, thanks to the precision of the orbital period, has a formal validity of 300 years. This ephemeris is suitable for future investigations of the orbital period changes.
\item The constant luminosity without outbursts and the absence of orbital humps suggest a nova-like classification.
\item Due to deep eclipses, \astrobj{J0130} is suitable to study the accretion disc structure using eclipse mapping methods.
\item Constraints on the inclination obtained from the eclipse will allow phasing the spectroscopic data and determining of the stellar masses with radial velocity measurements.

\end{enumerate}

\begin{flushleft}
{\bf{Acknowledgements}}
\end{flushleft}

\vspace{0.3cm}

This work was supported in part by the Ministry of Education and Science (the basic part of the State assignment, RK no.~AAAA-A17-117030310283-7) and by the Act 211 of the Government of the Russian Federation, agreement no.~02.A03.21.0006. This research has made use of the SIMBAD database, the NASA Astrophysics Data System (ADS), the International Variable Star Index (VSX) database and the VizeR catalogue access tool. The SIMBAD database is operated at CDS, Strasbourg, France. The VSX database is operated at AAVSO, Cambridge, Massachusetts, USA. The original description of the VizeR service was published in \citet{ochsenbein00}. This research has made use of Aladin sky atlas developed at CDS, Strasbourg Observatory, France. This work has made use of data from the European Space Agency (ESA) mission
{\it Gaia} ({https://www.cosmos.esa.int/gaia}), processed by the {\it Gaia}
Data Processing and Analysis Consortium (DPAC,
{https://www.cosmos.esa.int/web/gaia/dpac/consortium}). Funding for the DPAC
has been provided by national institutions, in particular the institutions
participating in the {\it Gaia} Multilateral Agreement.

\vspace{0.3cm}

\begin{flushleft}
{\bf{References}}
\end{flushleft}

\bibliographystyle{elsarticle-harv}

\vspace{0.5cm}


\begin{thebibliography}{}

\bibitem[Applegate(1992)]{applegate92}
Applegate,~J.~H., 1992. A Mechanism for Orbital Period Modulation in Close Binaries. ApJ 385, 621. https://doi.org/10.1086/170967.


\bibitem[Aungwerojwit et al.(2012)]{aungwerojwit12}
Aungwerojwit,~ A., G\"{a}nsicke,~B.~T., Wheatley,~ P.~ J., Pyrzas,~S., Staels,~B., Krajci,~T., Rodr\'{i}guez-Gil,~P., 2012. IPHAS~J062746.41+014811.3: A Deeply Eclipsing Intermediate Polar. ApJ 758, id. 79. https://doi.org/10.1088/0004-637X/758/2/79.


\bibitem[Baptista(2004)]{baptista04}
Baptista,~R., 2004. What can we learn from accretion disc eclipse mapping experiments? Astron. Nachr. 325, 181. https://doi.org/10.1002/asna.200310212.

\bibitem[Beuermann et al.(2011)]{beuermann11}
Beuermann,~K., Buhlmann,~J., Diese~J., et al., 2011. The giant planet orbiting the cataclysmic binary DP Leonis. A\&A 526, id. A53. https://doi.org/10.1051/0004-6361/201015942.

\bibitem[Bruch(2014)]{bruch14}
Bruch,~A., 2014. Long-term photometry of the eclipsing dwarf nova V893 Scorpii. Orbital period, oscillations, and a possible giant planet. A\&A 566, id. A101. https://doi.org/10.1051/0004-6361/201423576.

\bibitem[Downes et al.(1986)]{downes86}
Downes,~R.~A., Mateo,~M., Szkody,~P., Jenner,~D.~C., Margon,~B., 1986. Discovery of a new short-period, eclipsing cataclysmic variable. ApJ 301, 240. https://doi.org/10.1086/163893.

\bibitem[Eastman et al.(2010)]{eastman10}
Eastman,~J.,  Siverd,~R., Gaudi,~ B.~S., 2010. Achieving better than 1 minute accuracy in the heliocentric and barycentric julian dates. PASP 122, 935. https://doi.org/10.1086/655938.


\bibitem[Gaia Collaboration, Prusti et al.(2016)]{gaia16}
Gaia Collaboration, Prusti,~T., de~Bruijne,~J.~H.~J., Brown,~A.~G.~A., et al., 2016. The Gaia mission. A\&A 595, id. A1. https://doi.org/10.1051/0004-6361/201629272.


\bibitem[Gaia Collaboration, Brown et al.(2018)]{gaia18}
Gaia Collaboration, Brown,~A.~G.~A., Vallenari,~A., Prusti,~T., et al., 2018. Gaia Data Release 2. Summary of the contents and survey properties. A\&A 616, id. A1. https://doi.org/10.1051/0004-6361/201833051.



\bibitem[Hellier(2001)]{hellier01}
Hellier,~C., 2001. Cataclysmic Variable Stars, Springer. 



\bibitem[Kozhevnikov(2002)]{kozhevnikov02}
Kozhevnikov,~V.~P., 2002. Advantages of multichannel photometers in observations of rapid stellar oscillations and planetary transits. In: Battrick,~B., Favata,~F., Roxburgh,~I.~W., Galadi,~D. (Eds.), Proceedings of the First Eddington Workshop on Stellar Structure and Habitable Planet Finding, 11 -- 15 June 2001, C\'{o}rdoba, Spain. ESA SP-485, Noordwijk: ESA Publications Division, p. 299.

\bibitem[Kozhevnikov(2014)]{kozhevnikov14}
Kozhevnikov,~V.~P., 2014. Detection of eclipses in the suspected V Sge star IPHAS~J025827.88+635234.9. ApSS 349, 361. https://doi.org/10.1007/s10509-013-1648-2.

\bibitem[Kozhevnikov(2018)]{kozhevnikov18}
Kozhevnikov,~V.~P., 2018. Discovery of deep eclipses in the cataclysmic variable IPHAS J051814.33+294113.0.  ApSS 363, id. 130. https://doi.org/10.1007/s10509-018-3351-9.

\bibitem[Kozhevnikov and Zakharova(2000)]{kozhevnikoviz}
Kozhevnikov,~V.~P., Zakharova,~P.~E., 2000. The two-star photometer at Kourovka observatory: design and noise analysis. In: Garzon,~F., Eiroa,~C., de~ Winter,~D., Mahoney,~T.~J. (Eds.), ASP Conf. Ser. Vol. 219, Disks, Planetesimals and  Planets. Astron. Soc. Pac., San Francisco, p. 381. 

\bibitem[la~Dous(1994)]{ladous94}
la~Dous,~C., 1994. Observations and theory of cataclysmic variables: on progress and problems in understanding dwarf novae and nova-like stars. Space Sci. Rev. 67, 1. https://doi.org/10.1007/BF00750527.


\bibitem[Ma\'{i}z Apell\'{a}niz and Weiler (2018)]{maiz18}
Ma\'{i}z Apell\'{a}niz,~J., Weiler,~M., 2018. Reanalysis of the Gaia Data Release 2 photometric sensitivity curves using HST/STIS spectrophotometry. A\&A 619, id. A180. https://doi.org/10.1051/0004-6361/201834051.

\bibitem[Ochsenbein et al.(2000)]{ochsenbein00}
Ochsenbein,~ F., Bauer,~P., Marcout, J., 2000. The VizieR database of astronomical catalogues. A\&AS 143, 23. https://doi.org/10.1051/aas:2000169.

\bibitem[Ritter and Kolb(2003)]{ritter03}
Ritter,~H., Kolb,~ U., 2003. Catalogue of cataclysmic binaries, low-mass X-ray binaries and related objects (Seventh edition). A\&A 404, 301 (update RKcat
7.24, 2017). https://doi.org/10.1051/0004-6361:20030330.

\bibitem[Robinson(1992)]{robinson92}
Robinson,~E.~L., 1992. On the Reliability of White Dwarf Radial Velocity Curves determined from Emission-Line Velocities. In: Vogt,~N. (Ed.), ASP Conf. Ser. Vol. 29, Vi\~na Del Mar Workshop on Cataclysmic Variable Stars. Astron. Soc. Pac., San Francisco, p. 3. 

\bibitem[Rubenstein et al.(1991)]{rubenstein91}
Rubenstein,~E.~P., Patterson,~J., Africano,~ J.~ l., 1991. The orbital period changes of UX Ursae Majoris. PASP, 103, 258. https://doi.org/10.1086/132945.

\bibitem[Schwarzenberg-Czerny(1989)]{schwarzenberg89}
Schwarzenberg-Czerny,~A., 1989. On the advantage of using analysis of variance for period search. MNRAS 241, 153. https://doi.org/10.1093/mnras/241.2.153.

\bibitem[Schwarzenberg-Czerny(1991)]{schwarzenberg91}
Schwarzenberg-Czerny,~A., 1991. Accuracy of period determination. MNRAS 253, 198. https://doi.org/10.1093/mnras/253.2.198.

\bibitem[Schwarzenberg-Czerny(1998)]{schwarzenberg98}
Schwarzenberg-Czerny,~A., 1998. Period Search in Large Datasets. Baltic Astron. 7, 43. https://doi.org/10.1515/astro-1998-0109.


\bibitem[Szkody and Howell(1993)]{szkody93}
Szkody,~P., Howell,~S.~B., 1993. A spectroscopic study of DV Ursae Majoris (US 943), AY Piscium (PG 0134+070), and V503 Cygni. ApJ 403, 743. https://doi.org/10.1086/172245.

\bibitem[Warner (1995)]{warner95}
Warner,~B., 1995. Cataclysmic Variable Stars. Cambridge Astrophys. Ser., vol. 28. Cambridge University Press, Cambridge. 

\bibitem[Witham et al.(2007)]{witham07}
Witham,~A.~R., Knigge,~C.,  Aungwerojwit,~A., Drew,~J.~E., G\"{a}nsicke,~B.~T., Greimel,~R., Groot,~P.~J., Roelofs,~G.~H.~A., Steeghs,~D., Woudt,~P.~A., 2007. Newly discovered cataclysmic variables from the INT/WFC photometric H$\alpha$ survey of the northern Galactic plane. MNRAS 382, 1158. https://doi.org/10.1111/j.1365-2966.2007.12426.x.

\bibitem[Zorotovic et al.(2011)]{zorotovic11}
Zorotovic,~M., Schreiber,~M.~R., G\"{a}nsicke,~B.~T., 2011. Post common envelope binaries from SDSS. XI. The white dwarf mass distributions of CVs and pre-CVs. A\&A 536, id. A42. https://doi.org/10.1051/0004-6361/201116626.


\end{thebibliography}

\end{document}